\documentclass[preprint,showpacs,showkeys,amsmath,amssymb,aps,doublespace]{revtex4}
\usepackage[dvips]{graphicx}
\usepackage{amsmath}
\usepackage{amsfonts}
\usepackage{amssymb}

\begin{document}

\title{Improving Wang-Landau sampling with adaptive windows}

\author{A. G. Cunha-Netto\footnote{e-mail address: agcnetto@fisica.ufmg.br}$^{1,2}$,
A. A. Caparica$^1$,
Shan-Ho Tsai$^3$, Ronald Dickman$^2$ and D. P. Landau$^3$}
\affiliation{$^1$Instituto de F\'{\i}sica, Universidade Federal de Goi\'{a}s, C.P. 131, 74001-970,
Goi\^{a}nia, Brazil}
\affiliation{$^2$Departamento de F\'{\i}sica, Instituto de Ci\^{e}ncias Exatas,
Universidade Federal de Minas Gerais, C.P.702, 30123-970 Belo Horizonte, Minas Gerais, Brazil}
\affiliation{$^3$Center for Simulational Physics, University of Georgia, Athens, Georgia, 30602 USA}

\begin{abstract}
Wang-Landau sampling (WLS) of large systems requires dividing the energy
range into ``windows" and joining the results of simulations in each
window.  The resulting density of states (and associated thermodynamic functions)
are shown to suffer from boundary effects in simulations of lattice polymers
and the five-state Potts model.  Here, we
implement WLS using {\it adaptive} windows.
Instead of defining fixed
energy windows (or windows in the energy-magnetization plane for the Potts model), the boundary
positions depend on the set of energy values on which
the histogram is flat at a given stage of the simulation. Shifting the
windows each time the modification factor $f$ is reduced, we eliminate
border effects that arise in simulations using fixed windows.
Adaptive windows extend significantly the range of system sizes that may
be studied reliably using WLS.
\end{abstract}

\pacs{05.10.Ln, 64.60.Cn, 64.60.De}

\keywords{Monte Carlo simulation, Wang-Landau sampling, lattice polymer, Potts model}

\maketitle

In recent years Wang-Landau sampling (WLS) \cite{landau,landaupre,wanglandau_other,wang_landau_ajp} has
become an important
algorithm for Monte Carlo (MC) simulations, and is being applied to a vast array of models
in statistical physics and beyond\cite{wang_landau_ajp}.
WLS uses a random walk in energy ($E$) space to estimate the density
of states, $g(E)$; to sample well the full range of energies, the walk is adjusted to spend
approximately equal time intervals at each energy value.
The estimate for $g(E)$ is refined
successively, in a series of random walks.  WLS
has been used, for example, to simulate polymers \cite{3d,jain,rampf_binder_paul,parsons_williams} and
proteins \cite{rathore,wuest_landau} and to calculate the joint density of states (JDOS)
of two or more variables\cite{zhouetal}, e.g., $g(E,M)$ of magnetic systems
($M$ is the magnetization).

For models with a complex energy landscape, WLS permits simulation of larger
systems than can be studied using conventional MC approaches.
Because sampling
the full range of energies of a large system is not viable using a
single walk, one divides the
energy range into slightly overlapping subintervals (``windows") and samples
each separately.
The density of states $g(E)$ for the full energy range is then obtained
via a matching procedure that consists of multiplying
$g(E)$ in each window by a factor to force continuity of this
function at the borders.
(This yields $g(E)$ to within an overall multiplicative factor, permitting
evaluation of canonical averages. If the absolute density of states
is needed, the normalization must use some reference value for which the density of states
is known exactly, for example, the ground state degeneracy.)

The question naturally arises whether sampling restricted to windows yields reliable
estimates for $g(E)$, near the boundary energy values.
Wang and Landau suggested that boundary effects would
be negligible if adjacent windows were defined with a suitably large overlap \cite{landaupre}.
On the other hand, Schulz \textit{et al.}\cite{schulz} found it advantageous to
introduce an additional rule, namely, whenever a configuration is rejected
because its energy is greater than the maximum value, $E_>$, of a given window,
one should update $g(E)$ for the current energy value.
These prescriptions are effective for Ising models but do not eliminate boundary effects
in all instances, e.g., in simulations of polymers \cite{preprint} and systems where
the JDOS is required  \cite{shan}.

Here we present an implementation of WLS that eliminates border effects.
For the sake of clarity, we consider two examples with severe border problems: a lattice polymer
with attractive interactions between nonbonded nearest-neighbor
monomers \cite{baumgartner,3d,douglas_ishinabe}, simulated via reptation\cite{landau_binder},
and the calculation of $g(E,M)$ in the five-state Potts model \cite{fywu} on the square lattice.
(We note that while the reptation method is not suitable for sampling the most compact
configurations, this limitation does not affect the conclusions presented here.)

\begin{figure}[!ht]
\includegraphics[clip,angle=0,width=1.0\hsize]{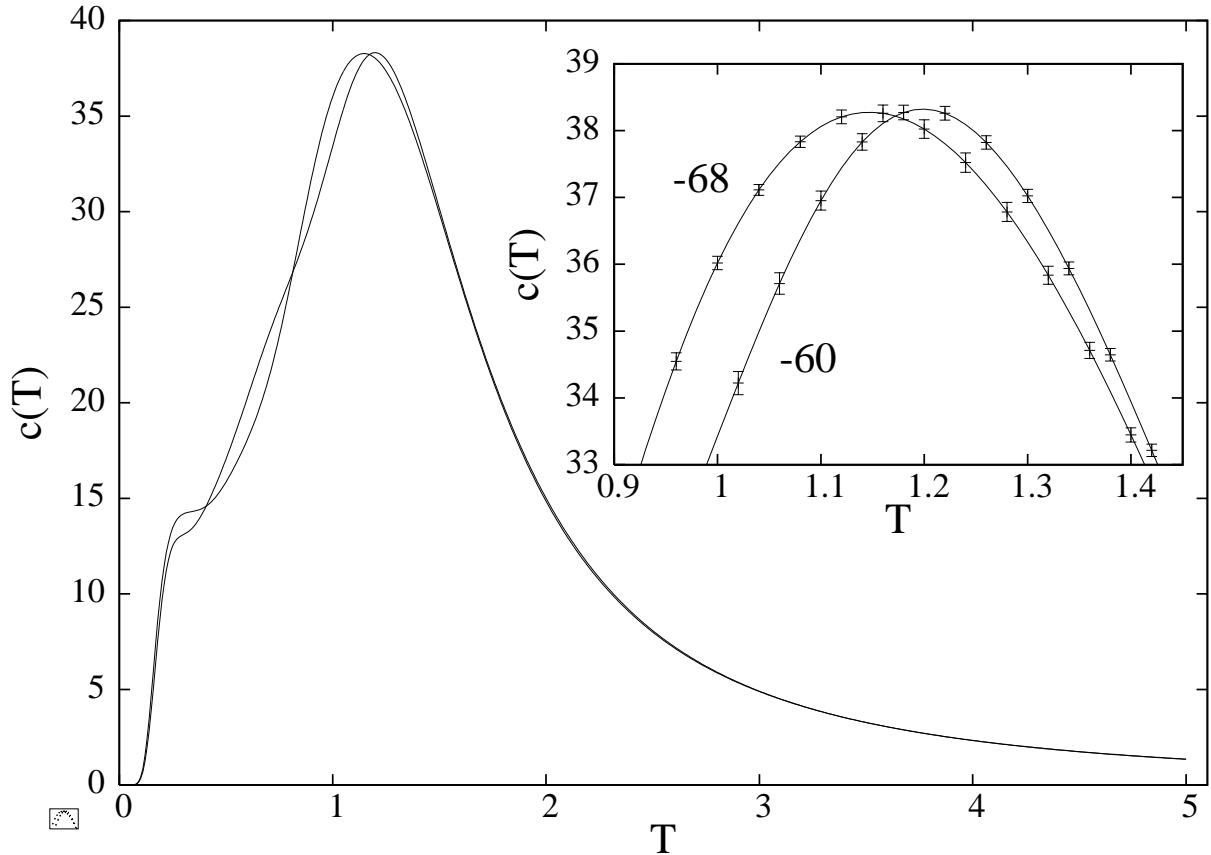}
\caption{Specific heat of a polymer of $N=100$ monomers on the square lattice,
using two windows and border energies $E_b$ = -68 and -60.
The inset shows the neighborhood of the maximum. Average values and uncertainties
are calculated using ten independent runs.}
\label{fig:border}
\end{figure}

To begin, we document the border effects that arise in fixed-window simulations
of polymers, focusing on the specific heat $c(T)$ (calculated as usual from
the variance of the energy). The polymer chain is modeled as a self-avoiding walk
on the square lattice.  The energy is $E=-N_{nb}$, with $N_{nb}$ the number of nearest-neighbor
contacts between nonconsecutive monomers. Our simulations sample the entire energy range,
from the ground state $E_{min}$ to $E_{max}=0$.
We encounter strong border effects, even if, following the suggestion of
Ref. \cite{schulz}, we use three overlapping levels at the border(s).
Varying the border energy $E_b$, we find that the temperature $T_m$ at which $c(T)$
takes its maximum varies in an analogous manner (a smaller $E_b$ corresponds to a smaller $T_m$),
signaling a systematic error (see Fig. \ref{fig:border}). Evidently WLS using fixed
windows yields distorted results for $g(E)$ in this system.  (We stress that such
a distortion does not arise in WLS of the 2d Ising model \cite{landaupre}, for which
simulations using fixed
windows have been performed on lattices of up to $256\times256$ sites.)

We shall now describe a procedure that eliminates the boundary effects
described above. Recall first that in WLS, the simulation begins with an arbitrary configuration,
and with $g(E)=1$ for all values of energy $E$. A random walk
in energy space is realized by generating trial configurations and accepting them
with probability $p(E_1 \rightarrow E_2)= \min(g(E_1)/g(E_2), 1)$,
where $E_1$ is the current energy value and $E_2$ the energy of the
trial configuration ${\cal C}_2$. If ${\cal C}_2$ is accepted,
we update $g(E_2) \rightarrow f\times g(E_2)$ and $H(E_2) \rightarrow
H(E_2)+1$; otherwise, $g(E_1) \rightarrow f\times g(E_1)$ and $H(E_1) \rightarrow
H(E_1)+1$, where the histogram $H(E)$ records the number of visits to energy $E$, and $f$ is the
modification factor, initially set to $e = 2.71828 \ldots$. When $H(E)$
is sufficiently {\it flat}, it is reset to zero and a new random walk is initiated,
with a smaller $f$, e.g., $f\rightarrow\sqrt{f}$, used to update $g(E)$. The simulation ends when $f$ is
sufficiently close to $1$, e.g., $f\approx 1+10^{-6}$.

The histogram is said to be flat if, for all energies
in the window of interest, $H(E) >\kappa \overline{H}$,
where the overline denotes an average over energies.
(Typically, $\kappa = 0.8$, as used here.)
In WLS with fixed windows, this stopping criterion is applied in each
window.  In our method, by contrast, we determine the range over which the histogram
has attained the desired degree of uniformity at various stages of the simulation.

\begin{figure}[!htb]
\centering
\includegraphics[clip,angle=0,width=1.0\hsize]{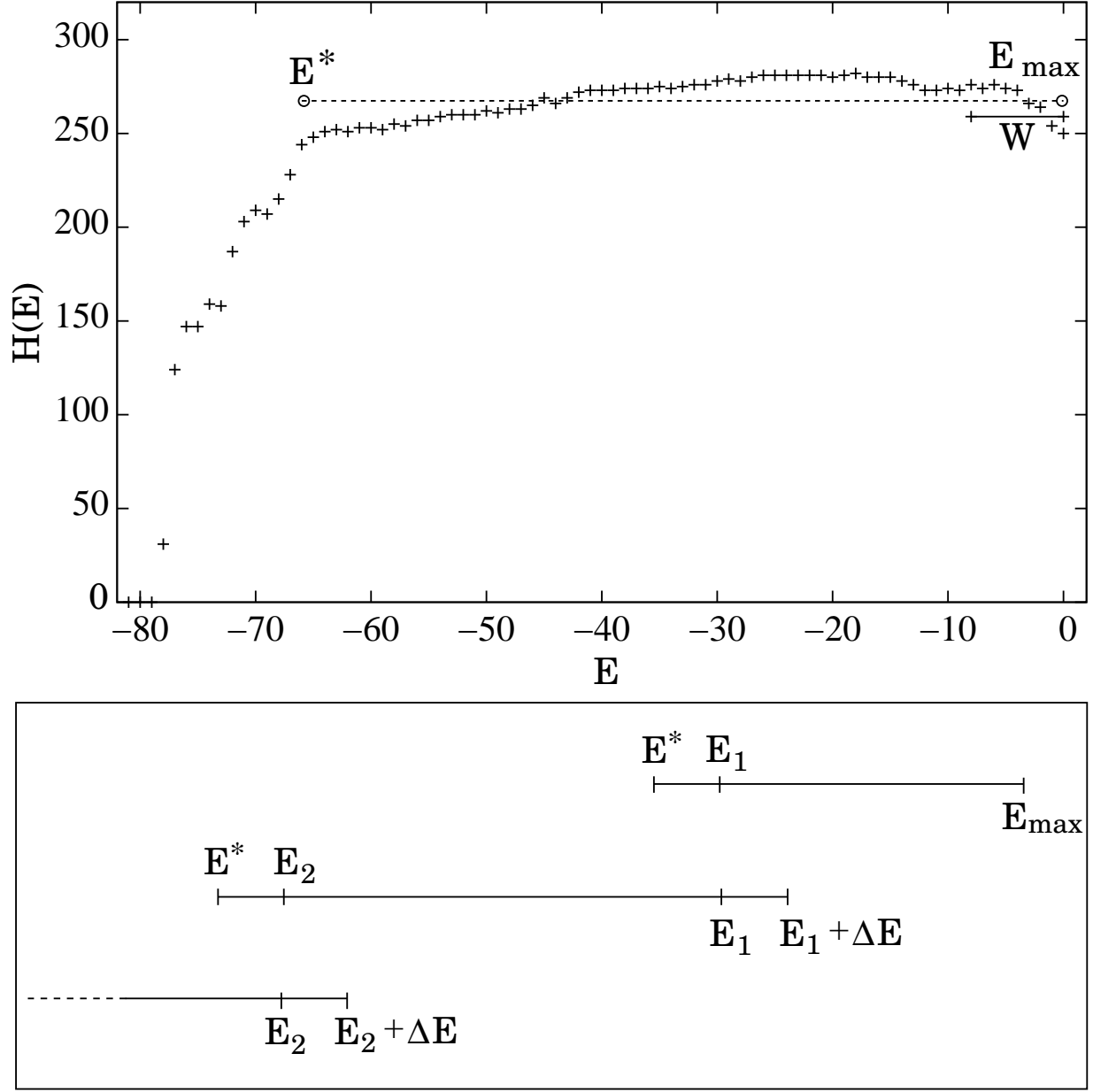}
\caption{Lower panel: schematic of adaptive windows; the value of $E_{max}$ depends on the
model, while $\Delta E$ is chosen to ensure sufficient overlap.  $E^*$ and $E_1$,
$E_2$, etc., are determined during the simulation not fixed beforehand.  Upper panel:
example of a histogram and associated window in
the lattice polymer simulation.}
\label{fig:schem}
\end{figure}

Initially, we allow the WLS random walk to visit all possible energies
$E_{min} \leq E \leq E_{max}$, accumulating the histogram in the usual manner.
After a certain number $N_1$ of Monte Carlo steps (in practice we use $N_1=10^4$),
we check if the histogram, restricted to the interval $E_w \leq E \leq E_{max}$, is flat.
(Here $E_w = E_{max} - W$, with $W$ defining the minimum acceptable window size.)
If it is not flat, we perform an additional $N_1$ Monte Carlo steps,
and check again, repeating until the histogram is flat on the minimal window.  Once
this condition is satisfied, we check whether the existing histogram is
in fact flat on a larger window.  (This is done by including the energy level
just below $E_w$ in $\overline{H}$ and checking if the flatness criterion,
$H(E) >\kappa \overline{H}$, holds in the enlarged window.  In this manner,
adding levels one by one, we identify the largest window over which flatness
is satisfied.)
Let $E^* \leq E_w$ be the smallest energy
such that the histogram, restricted to the interval $[E^*, E_{max}]$, is flat, and
let $E_1 = E^* + \Delta E$, where $\Delta E$ determines the overlap between adjacent
windows.
In the next stage of the simulation, we restrict the random walk to energies
$E_{min} \leq E \leq E_1 + \Delta E$.  After $N_1$ steps we check if the histogram is flat on the
interval $[E_1 + \Delta E - W, E_1 + \Delta E]$, and proceed as above,
until all possible energies have been included in a window with a flat
histogram (see Fig. \ref{fig:schem}). To avoid problems that might arise with very small windows,
the final window (with lower limit $E_{min}$) always has a width $\geq W$.
In the studies shown below we use $W=(E_{max}-E_{min})/10$
and $\Delta E=3$ for lattice polymers ($\Delta E=1$ or
$\Delta M=1$ for the Potts model).
Once all allowed energies have been sampled, we connect the functions $g(E)$ associated
with the various windows by imposing continuity at $E_1$, $E_2$,...,$E_n$.
Then the entire procedure is repeated using the next value of the modification factor $f$,
that is, $f \to \sqrt{f}$; the simulation ends once $f-1 < 10^{-6}$.

Note that at each iteration (using a different value of $f$), the borders
$E_j$ are chosen differently: repetition of a border value in successive iterations is prohibited.
This point is crucial to the functioning of our method; if the borders were fixed,
errors incurred at a given iteration (similar to those seen in Fig. \ref{fig:border}) would
accumulate, rather than being corrected in subsequent iterations.
(Here it is well to remember that WLS functions precisely because
errors in $g(E)$ incurred at a given value of $f$ tend to be corrected at subsequent stages.)

\begin{figure}[!htb]
\centering
\includegraphics[clip,angle=0,width=1.0\hsize]{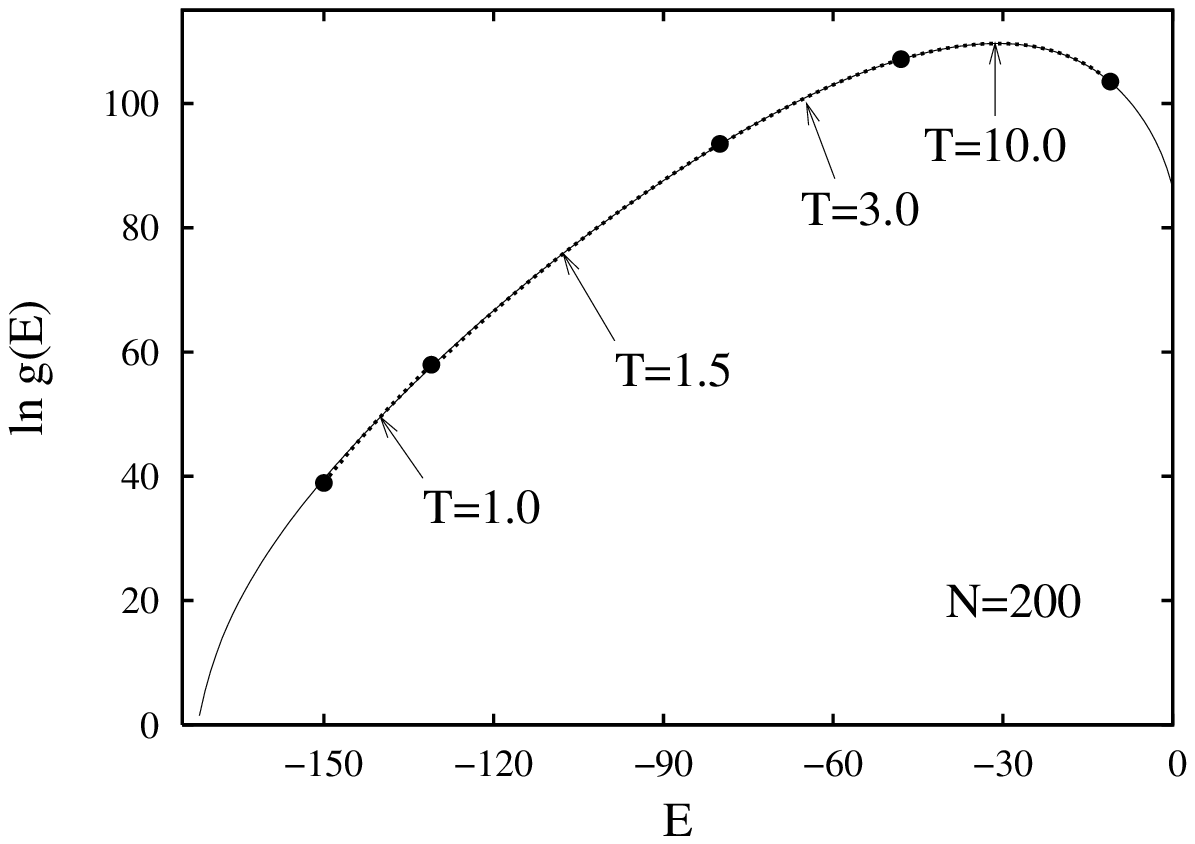}
\caption{Density of states $g(E)$ of two-dimensional polymers, $N=200$.
Solid line: adaptive-window sampling; dashed line: Metropolis sampling, using
the ratios $g(E)/g(E_{ref})$.  Circles denote the limits of the energy
ranges associated with each temperature in Metropolis sampling.}
\label{fig:dos200mt}
\end{figure}

It is important to stress that energy windows are not merely a device for
accelerating the simulation, but are in fact necessary in studies of large
systems; without them, WLS simply does not converge.  Thus, the largest
lattice polymer we are able to study without windows is $N=70$; using the
adaptive-window scheme, studies including the entire range of energies are
possible on chains of at least 300 monomers. To test the validity of our
method, we compare $g(E)$ (for $N=200$), given by our scheme with that
obtained using high-resolution Metropolis Monte Carlo\cite{landau_binder};
excellent agreement is found (see Fig. \ref{fig:dos200mt}). (Metropolis
simulations are performed at temperatures $T=1.0,1.5, 3.0$ and $10.0$,
using $10^8$ MC steps per study, using the RAN2 random number
generator \cite{numericalrecipes}.  In Metropolis MC, the expected number
$m(E)$ of visits to energy $E$ satisfies $m(E)/m(E_{ref}) =
e^{-(E-E_{ref})/k_B T} g(E)/g(E_{ref})$.  We use this relation to
determine the ratios $g(E)/g(E_{ref})$, by equating $m(E)$ to the actual
number of visits to energy $E$ during the simulation.  At each temperature,
a certain range of energies are well
sampled; for each temperature studied, we take the reference energy
$E_{ref}$ as the most visited value.  The resulting values for $g(E)$ are
normalized by equating $g(E_{ref})$ to the corresponding value obtained
via WLS.)

We turn now to the
$Q$-state Potts model\cite{fywu}, with Hamiltonian
\begin{equation}
{\cal H}=-J \sum_{\langle ij\rangle}\delta_{\sigma_i,\sigma_j} -h \sum_i\delta_{\sigma_i,1}
\label{eqn:HPotts}
\end{equation}
where $\sigma_i=1,2,..., Q$; $\langle ij\rangle$ denotes pairs of nearest-neighbor spins,
$J>0$ is a ferromagnetic coupling, and $h$ is an external field that couples to one of the states.
(Our units are such that $J/k_B=1$.)
We use WLS with a 2d random walk
to determine the JDOS $g(E,M)$. [Here
$E = -\sum_{\langle ij\rangle}\delta_{\sigma_i,\sigma_j}$
and $M=  \sum_i\delta_{\sigma_i,1}$.  Using $g(E,M)$,
thermodynamic properties may be obtained for arbitrary $(T,h)$.]
We study the five-state model on square lattices of 32 $\times$ 32 sites
with periodic boundaries;
we use the R1279 shift register random number generator \cite{landau_binder}.

\begin{figure}[ht]
\includegraphics[clip,angle=0,width=1.0\hsize]{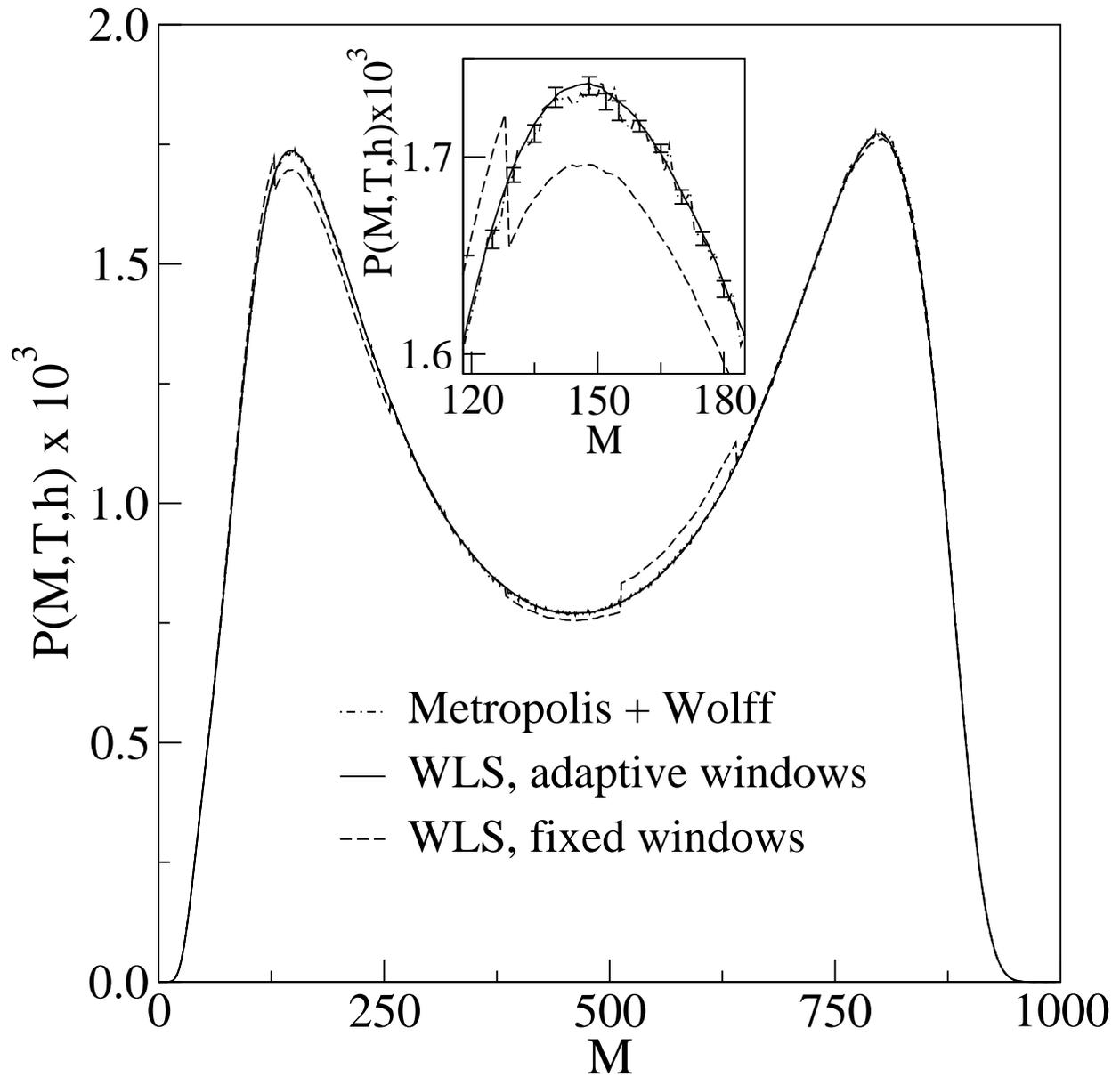}
\caption{Magnetization probability distribution $P(M,T,h)$ for
$T=0.86177$ and $h=0.005$ computed from $g(E,M)$ obtained via WLS
with fixed windows using case (ii) above (dashed line), adaptive windows
(solid line), and from a hybrid Monte Carlo (MC) method (dot-dashed line).
In the inset only a few typical error bars are shown for the hybrid MC result
(error bars for the WLS with adaptive windows are slightly smaller).}
\label{fig:probMQ5}
\end{figure}

\begin{figure}[ht]
\includegraphics[clip,angle=0,width=0.95\hsize]{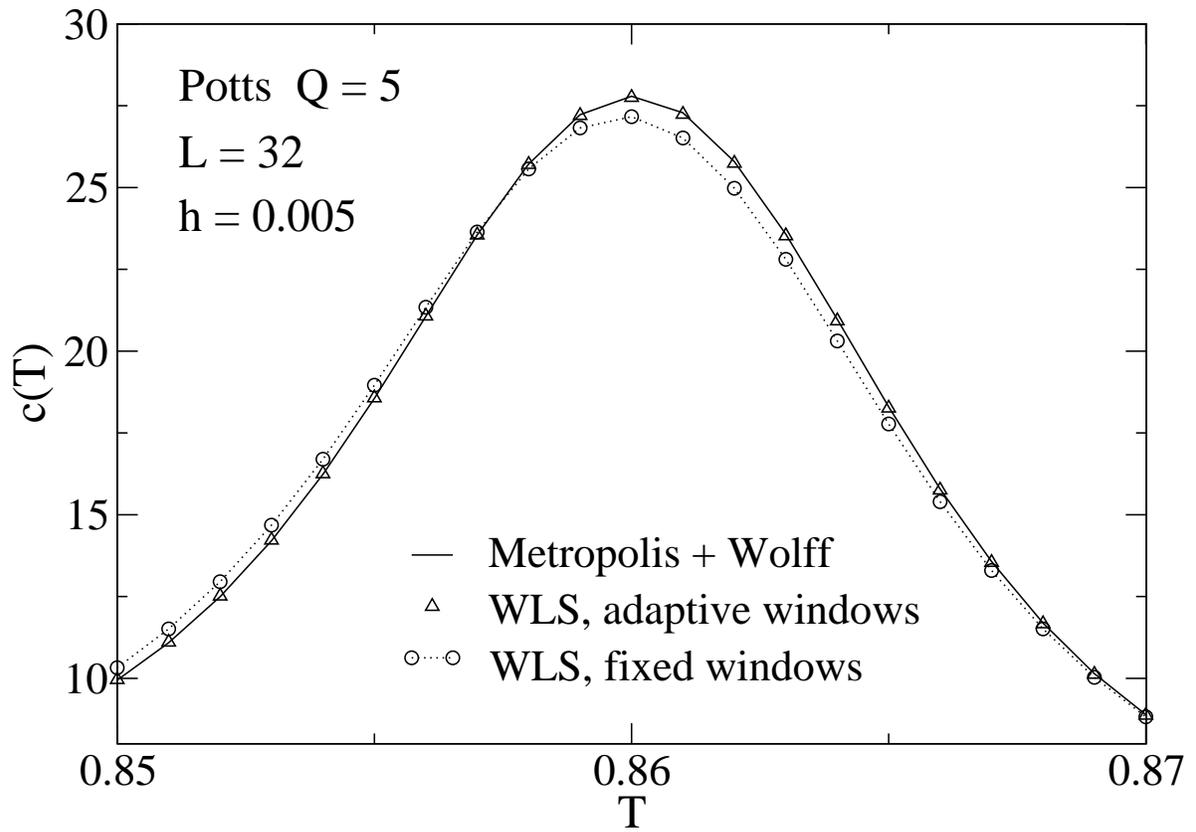}
\caption{Specific heat versus temperature
for $h=0.005$, obtained with different sampling
methods. Error bars are smaller than the symbol sizes.}
\label{fig:cvpotts}
\end{figure}

We tried to parallelize the 2d WLS by
performing independent random walks on overlapping $(E,M)$ regions.
We divided the parameter space into strips with: (i) restricted
values of $E$ and unrestricted $M$; (ii) restricted $M$ and
unrestricted $E$; and (iii) rectangles with both $E$ and $M$ restricted. The
independent random walks were carried out in parallel and the
densities of states from adjacent regions were
normalized at a single common $(E,M)$ point,
or by minimizing the least-square distance between
an overlapping $(E,M)$ region. We
found, however, that the resulting density of states
exhibits some small discontinuities which affect the
probability distributions of energy and magnetization, as
illustrated by the dashed line in Fig.\ref{fig:probMQ5} for $P(M,T,h)$
for the case (ii) above. This curve
was obtained with one simulation (hence
no error bars are displayed); the temperature is taken such that
$P(M,T,h)$ has two peaks of approximately
equal height. When we average $P(M,T,h)$ from multiple independent
simulations the jumps are somewhat smoothed out; nevertheless, the
thermodynamic quantities obtained with these implementations of parallel
WLS suffer from a small systematic error, as illustrated in
Fig.\ref{fig:cvpotts} for the specific heat.

The discontinuities in the probability distributions, and the
systematic error in thermodynamic quantities arise because
the slope of the JDOS is discontinuous at the borders between neighboring
fixed windows. These problems can be partly circumvented
by using a larger overlap region,
so that the JDOS used in the calculation of thermodynamic properties
does not come from the immediate vicinity of any border. The disadvantage of such an approach is that the
overlap between adjacent windows has to be quite large.  There is,
moreover, no guarantee that these regions will be sampled
properly. Using adaptive windows, by contrast,
the slopes of the JDOS at the borders of neighboring
windows are equal to within numerical uncertainty.
(In WLS with adaptive windows, we normalize the JDOS by applying a
least-square-difference criterion along the line of overlap between
neighboring sampling windows.)

The magnetization probability distribution computed from $g(E,M)$
(obtained from an average over ten independent runs using WLS with adaptive windows),
is also shown in Figure \ref{fig:probMQ5} (solid line).
This result is in good agreement with the distribution obtained using
a hybrid Monte Carlo method \cite{landau_binder} with $10^8$ MC steps.
Here one MC step comprises
$4L^2$ single spin-change trials with the Metropolis algorithm and $6$
Wolff cluster updates.
(In the main graph of Fig.\ref{fig:probMQ5} the result from adaptive-window WLS
is almost indistinguishable from that using hybrid MC.)
The specific heat obtained with WLS using
adaptive windows is also in excellent agreement with the result of
the hybrid Monte Carlo method, as shown in Fig.\ref{fig:cvpotts}.
Although we illustrate the effectiveness of WLS with adaptive windows
for $L=32$, a relatively small lattice size, the reader should recall
that we are performing \textit{two-dimensional} random walks; therefore,
the parameter space is much larger than for a one-dimensional random
walk using the same system size. The resulting JDOS allows us to obtain
thermodynamic quantities for the entire $(T,h)$ space. We note
in passing that because the 5-state Potts model has a very weak first-order
phase transition, the hybrid Monte Carlo method, employed here to test
our new method, can be used to sample equilibrium states; however, as
$Q$ increases, it becomes forbiddingly difficult for this method to
equilibrate the system using currently available computational resources.
In contrast, the WLS method works well even in the presence of strong
first-order phase transitions and, when combined with the adaptive windows
method presented here, can be used to simulate quite large systems.

In summary, we show how WLS, arguably the most
promising method presently available for simulating complex spin and fluid models,
may be applied reliably to large systems without border effects.  For the models
considered here, such effects are strong and effectively prohibit the
study of large systems using fixed windows.  In our method, errors that may arise near
the border of a given window are corrected in subsequent stages, in which the border positions
are shifted.  We expect this improvement of WLS to find broad application in
studies of polymers, spin systems with complex phase diagrams,
and complex fluids.

We are grateful to CAPES, CNPq, FUNAPE-UFG, Fapemig (Brazil),
and NSF grant number DMR-0341874 (USA) for financial support.

\end{document}